\documentstyle[12pt]{article}

\title{Hubble's Nobel Prize }

\author{D.S.L. Soares \\ Departamento de F\'{\i}sica, ICEx, UFMG 
--- C.P. 702 \\ 30161-970,  Belo Horizonte --- Brazil} 

\date{\today\\~ \\{\small [The Journal of the Royal 
Astronomical Society of Canada, vol. 95 , 10 (2001)]}}

\begin{document}

\maketitle

\begin{abstract}
Astronomy is not in the list of natural sciences aimed at by the Nobel 
awards.
In spite of that, there were, throughout the 1930s until the early 1950s,
effective moves by important scientists to distinguish Hubble with
the Prize. A short report on these attempts is made as well as 
speculation on what would be the citation for the prize in view of the broad
range of Hubble's scientific achievements. Within this context, the
opportunity is also taken for publicizing the Crafoord Prize which does
consider astronomy.
\end{abstract}

\section{Introduction}
The astronomer Edwin Powell Hubble never won the Nobel Prize because he
died unexpectedly, at the age of almost 64, on September 28, 1953, due to
a cerebral thrombosis. His long time family physician assured (!) his wife,
Grace Hubble, that the death had been ``instantaneous and without pain''
(Christianson 1995 --- hereafter CHR).

Is there a Nobel Prize for Astronomy? No, there is not! Hubble
would have been the first one to brake the old tradition and to change the
statutes of the award, as I show below. In the light of this
possibility, one might also wonder what Hubble would win the
Prize for, given his many scientific achievements.

But before that, in the next section, I describe another first rank
award series, also under the auspices of the Royal Swedish Academy of
Sciences, namely, the Crafoord Prize. This Prize includes astronomy in the
list of awards. In the final section, I comment on the story behind the
Nobel award to Hubble and speculate on the choice that would be made by
the Nobel Committee, from among his many fundamental investigations in
astronomy, as the Prize statement.

\section {The Crafoord Prize}
The Anna-Greta Holger Crafoord Fund was established in 1980 to promote
basic scientific research through yearly donations to the Royal Swedish
Academy of Sciences. Holger Crafoord (1908-1982) was very active 
in Swedish industry. From 1964 and on, he developed and manufactured
the artificial kidney, a sort of biological dialyzer that would become
of vital importance in the world. His company also developed a series
of medical instruments that contributed to earn him an enormous fortune.

In 1976 he became an honorary doctor of medicine at the University of Lund,
followed by his wife, Anna-Greta Crafoord (1914-1994), in 1987.

Specifically, the purpose of the Fund is to promote and award research in the
fields not covered by the Nobel Prizes in natural sciences, namely,
mathematics, geosciences, biosciences (with special emphasis in ecology
and rheumatoid arthritis), and astronomy. The awards follow a closed cycle
based on the annual sequence:
\begin{enumerate}
\item mathematics,
\item geosciences,
\item biosciences,
\item astronomy,
\item geosciences,
\item biosciences,
\item mathematics.
\end{enumerate}
The first Crafoord Prize was awarded in 1982 to V.I. Arnold, from 
Moscow State University, for his contribution to the theory of non-linear
differential equations. The Prize amounts to \$500,000 US, a gold medal, 
and a diploma.

The Crafford Prize is thus every six years assigned to astronomy. The first
recipient was Lyman Spitzer, Jr. (1985), then Allan R. Sandage (1991) and in
1997 there were two winners, Fred Hoyle and Edwin E. Salpeter.

More information on the Crafoord and Nobel Prizes are found in the
electronic pages of the Royal Swedish Academy of Sciences at the address 
{\tt www.kva.se/eng/pg/prizes/index.asp}.

\section {Hubble's Nobel}
As early as the 1930s (CHR), Fred Hoyle, a frequent guest of the
Hubbles at Pasadena, informed Hubble that there was a move, known of in
England, by the Nobel Prize Committee in the direction of a legal amendment
to the award statutes to make it possible for Hubble to be honoured
with this major distinction in natural sciences. The very same rumor was
also heard from Nobel laureate Robert Millikan, Caltech's celebrated
physicist.

Of course, Hubble had already been awarded many distinctive prizes, 
the highest being the Barnard medal, a charge of the National Academy of
Sciences, which was granted to him in 1935 at Columbia University. The
medal was established in 1895 and is awarded once every five years. All
of Hubble's predecessors were Nobel laureates, among them Roentgen, 
Rutherford,
Einstein, Bohr and Heisenberg (CHR). In this case Hubble was also
distinguished by being both the first American and the first astronomer to
win the medal. The citation in his award was for his {\it ``important studies
of nebulae, particularly of the extragalactic nebulae which provide the
greatest contribution that has been made in recent years to our observational
knowledge of the large-scale behavior of the Universe''}.

By 1949 nothing had yet happened, but with the support of
the 5-metre Palomar telescope, already in operation, Hubble's work had gained
much publicity, which could have triggered a decision by the Nobel Committee
(CHR). The final word was soon given when Enrico Fermi and
Subrahmanyan Chandrasekhar joined their colleagues in the Committee
unanimously voting Hubble the 1953 Prize in physics (CHR). But
it was too late, Hubble's death came first. Incidentally,
one should recall that the Nobel prize is not awarded posthumously.

These are the facts. Let us now speculate on an alternate universe, one
in which Hubble did survive that unfortunate September afternoon. 

The immediate question is what Hubble would be cited for in the
Nobel award. The conservative approach usually adopted by the Nobel
Committee is well known. The classical example is Einstein's. Awarded the
1921 prize, he was specially cited for his discovery of the law of the
photoelectric effect. It is needless to say that both the Special and General
Relativity Theories, which had already been put forward, were not explicitly 
mentioned.

Allan Sandage, a Crafford laureate and the greatest of Hubble's followers,
in a paper celebrating the centennial of the birth of Hubble (Sandage 1989),
enumerates Hubble's four central accomplishments undertaken from 1922 to 1936.
Sandage adds that any one of them guarantees Hubble a place in the history
of modern science. They are:
\begin{description}
\item{(a)} the morphological sequence of galaxy types, 
\item{(b)} the discovery of Cepheids in NGC6822, with parallel work in M31
and M33, settling decisively the question of the nature of galaxies,
\item{(c)} the determination of the homogeneity of the distribution of
galaxies, averaged over many solid angles, and
\item{(d)} the linear velocity-distance relation.
\end{description}
All of the four but one may be blurred with controversies, not only in
modern times but also, and certainly, in Hubble's time. Nowadays it is
recognized that the so-called Hubble  ``tuning-fork'' scheme in (a) applies
mainly to close and bright galaxies, that the homogeneity in (c) is broken by
the presence of enormous structures like ``walls'', ``streams'' and
``voids'' of galaxies, and, finally, even Hubble himself never clearly
advocated the idea of a velocity-distance relation in (d); rather he usually 
put it as a relation between spectral shift and distance, as is evident
throughout the classical book {\it The Realm of the Nebulae} (Hubble 1936).
In short, these three otherwise fundamental advances in modern
astronomy seem to collide with the spirit of Nobel citations, i.e., they
are not solid statements about nature.

This shortcoming is not the case, by any means, with item (b) above. Hubble's work
here is the end point of the great debate about the nature
of the nebulae. With the unambiguous determination of the distances to
the ``spiral nebulae'' to be much larger than the dimensions of our stellar
system, Hubble proved definitely that they are extragalactic, and 
as a result laid down the foundations of a new branch of research,
namely, extragalactic astronomy. The other three points mentioned by
Sandage are immediate consequences of this major realization, which were
soon recognized as such by the genius of Hubble. 

Thus, the Nobel prize to Edwin Powell Hubble goes for his {\it contribution
to the definitive understanding of the nature of the nebulae and for the
creation of a new era of scientific investigation}.

\section{References}
\begin{description}
\item{\bf Christianson G E} 1995 {\sl Edwin Hubble Mariner of the Nebulae},
The University of Chicago Press, Chicago
\item{\bf Hubble E P} 1936 {\sl The Realm of the Nebulae}, Yale University
Press, New Haven
\item{\bf Sandage A R} 1989 {\sl The Journal of the Royal Astronomical Society
of Canada} {\bf 83}, 351
\end{description}

\bigskip\bigskip\bigskip

{\noindent\it Acknowledgments ---} I wish to thank Dr. P.S.S. Guimar\~aes
for reading the manuscript and for useful suggestions.

\end{document}